\documentclass[10pt]{article}

\usepackage[utf8]{inputenc}
\usepackage[T1]{fontenc}
\usepackage{lmodern}
\usepackage[margin=0.75in]{geometry}
\usepackage{graphicx}
\usepackage{amsmath}
\usepackage{booktabs}
\usepackage{caption}
\usepackage{subcaption}
\usepackage{hyperref}
\usepackage{authblk}
\usepackage{xcolor}
\usepackage{array}
\usepackage{balance}
\usepackage{tikz,pgfplots,pgfplotstable}
\pgfplotsset{compat=1.18}
\usetikzlibrary{positioning,arrows.meta,fit,shadows.blur}
\usepackage{graphicx}
\usepackage{caption}
\usepackage{placeins}
\graphicspath{{fig/}}

\usepackage[numbers,sort&compress]{natbib}

\hypersetup{
    colorlinks=true,
    linkcolor=black,
    filecolor=magenta,      
    urlcolor=cyan,
    citecolor=black,
    pdftitle={Alignment Debt: The Hidden Work of Making AI Usable},
    pdfauthor={Cumi Oyemike}
}

 \usepackage{orcidlink}
 
\title{\textbf{Alignment Debt: The Hidden Work of Making AI Usable}}

\author[1]{Cumi Oyemike\thanks{Lead author and primary contributor.}~\orcidlink{0009-0007-1390-2192}}
\author[1]{Elizabeth Akpan}
\author[1]{Pierre Hervé-Berdys}

\affil[1]{YUX Design}

\date{October 2025}

\begin{document}
\maketitle
\begin{abstract}
Frontier LLMs are optimised around high‑resource assumptions about language, knowledge, devices, and connectivity. Whilst widely accessible, they often misfit conditions in the Global South. As a result, users must often perform additional work to make these systems usable. We term this \textit{alignment debt}: the user‑side burden that arises when AI systems fail to align with cultural, linguistic, infrastructural, or epistemic contexts. 

We develop and validate a four-part taxonomy of alignment debt through a survey of 411 AI users in Kenya and Nigeria. Among respondents measurable on this taxonomy ($n=385$), prevalence is: Cultural and Linguistic (51.9\%), Infrastructural (43.1\%), Epistemic (33.8\%), and Interaction (14.0\%). Country comparisons show a divergence in Infrastructural and Interaction debt, challenging one‑size‑fits‑Africa assumptions. 

Alignment debt is associated with compensatory labour, but responses vary by debt type: users facing Epistemic challenges verify outputs at significantly higher rates (91.5\% vs.\ 80.8\%; $p = 0.037$), and verification intensity correlates with cumulative debt burden  (Spearman’s $\rho = 0.147$, $p = 0.004$). In contrast, Infrastructural and Interaction debts show weak or null associations with verification, indicating that some forms of misalignment cannot be resolved through verification alone. 

These findings show that fairness must be judged not only by model metrics but also by the burden imposed on users at the margins, compelling context-aware safeguards that alleviate alignment debt in Global South settings. The alignment debt framework provides an empirically grounded way to measure user burden, informing both design practice and emerging African AI governance efforts. 
\end{abstract}


\section{Introduction}

Globally deployed AI systems arrive as a single interface, while users bring many worlds. In many Global South contexts, this creates a fundamental misalignment between design assumptions and local realities, and the work of bridging that gap is routinely offloaded onto users themselves.  We name that offloading \textit{alignment debt}: a user‑centred construct for the burdens that arise when AI misfits cultural and linguistic practices, infrastructural conditions, epistemic needs, and interaction literacies. Rather than adding another system‑side fairness metric, alignment debt reframes fairness as a question of who pays, foregrounding user‑side costs and the distribution of compensatory work required to make systems usable. In doing so, it complements audits that document disparities in model outputs by turning to the lived experiences of those disparities, a dimension richly theorised but rarely measured in African AI deployments.

Algorithmic fairness research has shown that data and design choices embed social hierarchies into AI systems, yielding uneven performance and representation across demographic and linguistic groups. For language technologies, corpus skews and representational gaps leave many African languages and knowledge systems poorly modelled, with downstream effects on intelligibility, relevance, and trust. African AI scholarship has further shown that infrastructural constraints such as intermittent connectivity, device limitations, data costs, interact with cultural and linguistic misfit to shape how, and at what cost, users can access AI capability. Yet much of this literature remains system‑centric: it quantifies model disparities but not the user‑experienced burden of living with them day‑to‑day.

We advance alignment debt to fill this gap. Practically, we operationalise alignment debt conservatively through user-reported indicators of misalignment and per‑respondent accumulation (0-4), and examine verification as an associated coping behaviour rather than a proxy for trust. This approach allows us to quantify burden without assuming unmeasured costs (time, data, cognitive load, task loss), while enabling rigorous prevalence estimation and behavioural linkage in situ.

Our lens synthesises adjacent literature. From algorithmic fairness, we retain the imperative to detect disparate harms but shift emphasis from score parity to to the distribution of compensatory work required from users. From African and Global South HCI, we engage the insight that exported systems often encode Global North values and interaction norms, shifting articulation work onto users and producing patterned inequalities of effort. From African NLP, we build on evidence that language coverage and representational depth are decisive for usability and trust in multilingual societies, especially where code‑switching, dialectal variation and low‑resource languages are prevalent. Alignment debt integrates these insights into a measurable, user‑centred criterion for whether \textit{works} where it is used, for whom, and at what cost.

\subsection{Research Questions}

This study addresses three questions:

\textbf{RQ1 (Prevalence of Alignment Debt)}: What is the prevalence of different alignment debt types among AI users in Kenya and Nigeria?

\textbf{RQ2 (Cross-Country Variation)}: How do alignment debt patterns differ between Kenya and Nigeria?

\textbf{RQ3 (Verification Behaviour)}: How does experiencing alignment debt relate to verification behaviour? We examine whether debt intensity predicts compensatory labour, and is this association specific to epistemic challenges rather than universal?

\subsection{Contributions}

This work makes four contributions:

\textbf{Empirical.} We provide the first large-scale quantitative measurement of alignment debt in African AI use (\textit{N}=411), demonstrating that misalignment is not incidental but widely experienced across cultural/linguistic, infrastructural, epistemic, and interaction dimensions.

\textbf{Theoretical.} We extend user burden theory~\cite{Suh2016} by introducing four AI-specific debt categories and linking them to behavioural coping practices. Where fairness asks \textit{“Are outcomes equal?”}, alignment debt asks \textit{“Who pays to make the system usable?”}.

\textbf{Methodological.} We develop and validate a reusable survey instrument and taxonomy enabling cross-cultural measurement of alignment debt. This provides a foundation for comparative Global South AI research.

\textbf{Practical.} We translate our findings into actionable design and governance implications. These include culturally grounded data practices, bandwidth-adaptive interfaces, epistemic transparency supports, and user-burden indicators for emerging African AI strategies.

\section{Related Work}

\subsection{Algorithmic Bias in African Contexts}

Extensive research on algorithmic fairness has identified and documented bias across domains such as facial recognition~\cite{Buolamwini2018}, healthcare algorithms~\cite{Obermeyer2019}, search engines~\cite{Noble2018}, and even the language models we interact with daily~\cite{Bender2021,Bolukbasi2016}. Scholars have raised concerns about the application of Global North training data, evaluation metrics, deployment assumptions, and performance benchmarks to African contexts without any meaningful adaptation~\cite{Birhane2020,Mohamed2020}. These concerns are well founded, as Western-trained models frequently fail to account for local African linguistic patterns and cultural practices, let alone the infrastructural conditions on the ground~\cite{Nekoto2020,Orife2020}. 

Unsurprisingly, such misalignment results in interactional failures; for example, speech and NLP systems show performance gaps on ethnolinguistic and accented varieties, and code-switched inputs remain poorly handled across tasks~\cite{Koenecke2020,Winata2023}. Drawing on a collection of more than 67,000 audio clips recorded across thirteen African countries, the AfriSpeech-200 dataset exposes how current speech recognition systems continue to falter on African-accented English~\cite{Olatunji2023}. Performance parity with "standard" benchmarks remains elusive unless models are explicitly adapted to local accents. This reveals how deeply accent bias is baked into training pipelines from the start. Language models show the same pattern, however, fluency can mask misunderstanding. Text may read smoothly yet flatten or erase the cultural references that give meaning its depth, leaving users with outputs that sound confident but are contextually false~\cite{Sheng2019,Bender2021}.

Against this backdrop, the African Union's Continental AI Strategy explicitly warns that Western-biased data and teams can “perpetuate or amplify biases” in AI and mandates impact assessments~\cite{AfricanUnion2024}. Asiedu et al.~\cite{Asiedu2024} call for fairness frameworks specific to African health AI, arguing that Western fairness metrics inadequately capture African priorities. Other commentaries underscore the need for empirical research into how algorithmic harms unfold in everyday African contexts, and for accountability mechanisms that are responsive to local cultural, linguistic, and infrastructural contexts~\cite{Birhane2020,Pasipamire2024,Mahamadou2024}.

Yet these strategic and scholarly calls have rarely been matched with grounded evidence of misalignment in practice. Despite the warnings and aspirations for locally responsive AI, relatively few empirical studies have traced the downstream burdens that misfitting systems impose on users in African settings. Our work moves beyond a one-dimensional view of bias as a statistical artefact to frame it as a lived condition that places measurable burdens on users. Examining AI use in Kenya and Nigeria, we quantify these burdens and highlight the costs of technological misalignment that existing fairness metrics often overlook.

\subsection{User Burden Theory and Measurement}

Human-computer interaction research formalises user burden as the multidimensional cost of interacting with technical systems \cite{Suh2016}. The User Burden Scale by Suh et al.\ \cite{Suh2016} measures six burden dimensions: difficulty of use, physical effort, time burden, mental and emotional burden, privacy concerns, and financial costs. Their work established a strong correlation between total burden and technology abandonment \cite{Suh2016}. However, the User Burden Scale and related instruments were developed for general software evaluation and validated on online survey samples, and thus they do not capture AI-specific burdens that are salient in Global South contexts, including cultural and linguistic misfit, epistemic effort spent verifying outputs, and prompt-engineering difficulty.

From emerging research into the user experience of AI systems, new burdens are being identified \cite{Wischnewski2023,Simkute2025}. Trust calibration, that is aligning user trust with an AI system's actual reliability, imposes cognitive costs because users often must evaluate outputs in the absence of clear reliability indicators \cite{Wischnewski2023,Okamura2020}. Frustration arising from repetitive, low-quality generative outputs and from unpredictable model behaviour further adds to user burden \cite{Simkute2025,ZamfirescuPereira2023}. Studies of explainability and recommender interfaces show that users expend cognitive effort when they weigh whether to accept or reject algorithmic suggestions \cite{Lee2004,Papenmeier2019}.

These empirical studies are valuable, yet there is a persistent concern about the generalisability of findings: much HCI and trust research relies heavily on Western, Educated, Industrialised, Rich and Democratic samples, which limits how readily results translate to Global South settings \cite{Henrich2010,Seaborn2023}. We extend this line of work by introducing a four-category, AI-specific taxonomy that operationalises the burdens most relevant to African contexts.

\subsection{Cross-Cultural AI and Postcolonial Computing}

Culture influences how people ask questions, judge answers, and decide whether to trust a system. Current large language models do not enter this space as neutral observers, they lean toward Western cultural priors \cite{Tao2024,Naous2025}. Recent evaluations across many countries and languages suggest systematic value skews: for instance, Hofstede-style assessments show that model responses often track English-speaking and Protestant European populations, while persona-based prompting across 36 countries reveals uneven cultural value representation \cite{Tao2024,Kharchenko2024}. Likewise, benchmarks of everyday cultural knowledge indicate large performance differences across regions and languages, which hints at gaps in the cultural materials these systems internalise \cite{BLEND2024}. Cross-lingual interaction remains a practical source of friction: LLMs sometimes fail to respond in the requested language or switch languages unexpectedly, creating “language confusion” that pushes users to adapt \cite{Marchisio2024}.  In India, for instance, people cope by switching languages, simplifying queries, or abandoning features entirely \cite{Sambasivan2021}.

Postcolonial computing scholarship furnishes a theoretical grounding for why these patterns emerge. Human-computer interaction studies often replicate colonial power dynamics by treating Western users as the default and other users as edge cases \cite{Irani2010}. We see this in how systems designed originally for Western contexts are superficially adapted for global deployment, often through language translation alone \cite{Dourish2012}. Similarly, assumptions of constant connectivity, personal device ownership, and particular privacy norms tend to mirror Western middle-class experiences, so the costs of making systems usable in other settings are foisted on users rather than absorbed by designers \cite{Dourish2012}. 

These design defaults sit within a broader political economy. Global data flow is theorised as a form of data colonialism, where information extracted from the Global South fuels innovation in the North whilst decision-making power and economic benefit remain concentrated there \cite{Couldry2019}. Users in these regions remain underrepresented in the knowledge sources that feed many large-scale training corpora, resulting in informational geographies where African contexts are far less visible \cite{Graham2014}. Beyond language, Eurocentric training data can also perpetuate colonial patterns in cultural domains, marginalising African art through reductionist approaches and epistemic imbalance \cite{Bignotti2025}.

Attempts to repair these cultural misalignments are emerging. In education settings, for example, it has been suggested that prompt engineering might serve as a practical skill to help teachers and students steer models toward culturally meaningful content, though the argument is conceptual rather than empirically tested at scale \cite{Murungu2024}. Meanwhile, systematic reviews of prompt-based debiasing show that certain techniques can substantially mitigate cultural bias in model outputs, though the effectiveness varies widely and requires users to learn specialised interaction strategies \cite{Asseri2025}. 

\subsection{LLM Training Data Disparities}

Large language models reflect the composition of the corpora on which they are trained, and those corpora are unevenly distributed across languages and geographies. Most commercial large language models are trained primarily on English and Western-centric web data, a trend observable in the last model whose training composition was publicly disclosed \cite{Brown2020,Bender2021}. Subsequent proprietary models have withheld comparable disclosures, but performance patterns across multilingual benchmarks suggest similar linguistic asymmetries \cite{Alhanai2025,NakatumbaNabende2024}. Likewise, mapping studies of Wikipedia content have repeatedly shown that North America and Western Europe account for a large share of geolocated articles while many regions, including much of Africa, remain sparsely represented \cite{Graham2014,Graham2011}. A white paper on low-resource languages notes that many African languages are effectively excluded from the training of foundational models \cite{Pava2025}. Recent empirical work quantifies this more precisely: one study shows that multilingual models perform considerably worse on non-English inputs, and that fine-tuning with culturally relevant data improves performance by only a few percentage points but still leaves a substantial gap \cite{Alhanai2025}. Another study confirms that training-set composition produces epistemic gaps since the available corpora tend to reflect who publishes online and in which languages, not the full range of human knowledge \cite{Manvi2024}.

Such gaps have concrete downstream consequences. Work on multilingual and low-resource NLP documents a persistent scarcity of high-quality training and benchmark data for many African languages, and community efforts such as Masakhane and MasakhaNER explicitly treat this under-representation as an obstacle to usable systems \cite{Adelani2021}. Reviews and mapping efforts similarly note that, whilst Africa is home to over two thousand languages, the majority are severely under-resourced in current NLP datasets and model evaluations \cite{NakatumbaNabende2024}. As a result, queries about region-specific history, medical practice, or local institutions can return fluent but culturally skewed or incomplete responses, which then require local verification \cite{Talat2022}.

\subsection{Information Verification and Trust}

Calibrating trust in AI outputs imposes real cognitive and temporal costs on users. Trust calibration is a well-studied concern across human-computer interaction and human-AI interaction, with survey and experimental work documenting the risks of both over-reliance and under-reliance on automated suggestions \cite{Wischnewski2023,Okamura2020}. In everyday practice, people protect themselves from unreliable outputs through verification behaviours, such as cross-checking claims against trusted sources, consulting multiple models, or asking local experts \cite{ZamfirescuPereira2023,Lee2021}. Each of these strategies buys reliability but consumes scarce attention and often concrete resources, as studies show this calibration is cognitively and temporally demanding \cite{Afroogh2024}. Trust and its counterpart, scepticism, influence whether users accept an output without question or reject it entirely \cite{Durán2025}.  

Most research on trust and verification in AI has focused on high-stakes domains such as clinical decision support or legal advice, where error costs are obvious and measurement is relatively direct \cite{Mehrotra2024,Wischnewski2023,Jones2023,Turner2024}. Less attention has been paid to the cumulative costs of routine verification in general-purpose AI use, especially in contexts where training data gaps make outputs systemically less dependable. In low-resource environments, these costs are amplified by infrastructural constraints: intermittent connectivity, scarce local reference sources, limited access to paywalled journals or authoritative databases, and the high price of mobile data all increase the time and difficulty of verifying a model’s answer \cite{NakatumbaNabende2024,Coelho2025,Azaroual2024}. A large-scale survey of global AI attitudes found that in emerging economies, users are more likely to report worry about accuracy and data fairness compared to users in advanced economies \cite{Gillespie2023}.  

Building on this literature, our study treats verification as a coping behaviour and measures two facets: verification propensity (whether respondents report verifying model outputs at all) and verification intensity (the number of sources respondents report consulting when they verify). We then examine how these measures relate to alignment debt type.

\subsection{Synthesis}

Synthesising the literature, we see a clear pattern. Models trained and evaluated through Western lenses misfit everyday African use, and the burden of adaptation is routinely shifted onto users. Whist technical audits identify where these systems fail, and policy frameworks outline the governance risks, we still lack empirical evidence of the lived experience of these misalignments or what strategies users employ to compensate. To address that blind spot, we introduce \emph{alignment debt}: a user-centred lens for the burdens created when globally deployed AI fails to meet local cultural, linguistic, infrastructural, or epistemic conditions. The framework below specifies alignment debt, situates it within existing theory, and operationalises it into a taxonomy we can measure in situ. 

\section{Conceptual Framework}

\subsection{Definition and scope}

We use the term \textit{alignment debt} to refer to the user-side burden that accumulates when AI systems fail to align with the cultural, linguistic, infrastructural, epistemic, or interactional conditions in which they are used. Rather than framing misalignment as a property of the system alone, alignment debt foregrounds how users must compensate for these gaps through additional effort, adaptation, verification, or workaround strategies. The construct is descriptive and contextual: it captures the lived, ongoing labour required to make AI usable in practice, and recognises that this burden is unevenly distributed across settings and communities.

\subsection{Theoretical Foundations}

Our lens integrates three theoretical foundations. From postcolonial computing, we retain the critique that design often treats Western users as default and others as edge cases, externalising the work of contextual fit to those at the margins \cite{Irani2010,Dourish2012}. From data and knowledge geographies, we take the observation that public corpora and platform infrastructures unevenly represent African contexts, shaping what models learn and what they miss \cite{Graham2014,Couldry2019}. From human-computer interaction, we draw on user burden theory and adopt the idea that \emph{burden} is multidimensional and linked to abandonment, but we extend it with AI-specific forms that existing scales were not built to capture \cite{Suh2016}. The contribution here is to unify these perspectives around the question of \emph{who performs the compensatory labour when systems misfit their use}. In this study, we approach that question by measuring verification as one form of user-side burden and examining how it varies across different types of misalignment.

\subsection{Dimensions of alignment debt}

Four forms of alignment debt emerge consistently in African HCI and NLP research, and in the qualitative responses from our survey. These dimensions capture how users experience and compensate for misalignment in everyday interaction with AI systems.

\paragraph{Cultural and Linguistic debt.} 
Arises when systems fail to recognise or appropriately interpret users’ linguistic practices, communication styles, or cultural frames of reference. It appears in accent misrecognition, dialect flattening, code-switching breakdowns, and outputs that rely on Western social norms or cultural assumptions. It also includes cases where systems reproduce culturally biased stereotypes or adopt judgemental tones that do not align with local communicative norms. Users respond by rephrasing prompts, avoiding certain languages or dialects, or adopting a more “standard” or Western-aligned tone in order to be understood.

\paragraph{Infrastructural debt.} 
Occurs when systems are designed for environments with stable connectivity, low latency, and high device capability. In contexts where mobile data is expensive, network reliability is uneven, or devices are older, the cost of accessing AI becomes time-intensive and financially burdensome. In Kenya and Nigeria, respondents described waiting for pages to load, retrying stalled queries, or abandoning usage when the cost of data outweighed the benefit of the output.

\paragraph{Epistemic debt.} 
Reflects the burden placed on users when outputs are unreliable, locally irrelevant, or insufficiently evidenced. Models may hallucinate sources, present outdated or non-African reference points as defaults, or supply confident but contextually incorrect information. This compels users to verify, cross-check, or seek external confirmation. Epistemic debt therefore converts informational uncertainty into user labour.

\paragraph{Interaction debt.}
This arises when the interaction model of the system does not match how users structure tasks or communication. Users describe needing to learn specialised prompting strategies, re-establishing lost conversational context, or adapting workflows to suit the system rather than the task. Interaction debt compounds when cultural and linguistic misalignment is also present, amplifying cognitive load and frustration.

\subsection{Taxonomy development}

Alignment debt is operationalised as four observable categories that reflect distinct ways AI can misfit everyday use. These categories were derived from qualitative responses in the survey instrument and from recurrent patterns documented in African HCI and NLP research. Each category is conservatively coded as present or absent per respondent, and a cumulative index (0-4) reflects the total number of distinct misalignments experienced, without assuming a single latent factor.

\medskip
All four types can co-occur, but they are not assumed to reduce to a single construct. The taxonomy therefore treats alignment debt as a profile rather than a score, and cumulative burden as the count of distinct misalignments that a single user must manage in practice.
\begin{table}[ht]
\centering
\caption{Operationalisation of alignment debt categories from survey responses. Respondents could select multiple challenges. Each category is coded as present if any mapped item was selected. Privacy-only selections were excluded from taxonomy scoring.}
\label{tab:taxonomy_operationalisation}
\begin{tabular}{p{3.2cm} p{5.8cm} p{5.8cm}}
\toprule
\textbf{Debt category} & \textbf{Survey items mapped} & \textbf{Interpretation / observable indicators} \\
\midrule
Cultural and Linguistic debt &
Misunderstands my accent/language; Lacks cultural awareness; It is biased; Feels judgmental or rude &
Model requires users to adjust language, framing or tone; cultural references are flattened or misinterpreted. \\

Infrastructural debt &
Needs stronger internet signal; Takes too long to load answers &
System assumes connectivity or device capacity that does not hold; users experience latency, cost or access friction. \\

Epistemic debt &
Gives wrong answers; Shares misinformation; Doesn’t list its sources or share thought process &
User must verify correctness and provenance before acting on outputs. \\

Interaction debt &
I struggle with prompting or being understood by LLMs &
User performs orchestration work (rephrasing, scaffolding, reframing) to obtain intended output. \\
\midrule
\textbf{Excluded from taxonomy scoring} &
Shares my data without consent &
Privacy concerns are meaningful but analytically distinct from alignment debt. Respondents selecting only this item were excluded from taxonomy calculations. \\
\bottomrule
\end{tabular}
\end{table} 

\section{Methods}

This study uses a cross-sectional survey of AI users in Kenya and Nigeria. All participants provided informed consent before survey access. This section describes the survey design, participant recruitment, alignment debt measurement, data quality checks, and analytical approach.

 \subsection{Survey Design and Instrument Development}

The survey was administered using LOOKA, a multilingual, mobile-optimised survey platform designed for research in African settings. Taking into account demographic differences, the instrument was deployed in both English and French. LOOKA's infrastructure enabled efficient distribution through social media channels and professional networks.  The full questionnaire (42 items with conditional branching) is provided in Appendix A.

The instrument consisted of five sections. Section A captured demographic information and general AI usage frequency. Section B collected data on devices and connectivity conditions. Section C asked about commonly used AI tools, preferred interaction formats, usage contexts, and verification behaviours. Section D contained conditional blocks probing trust and reasoning across six use-case domains (personal, professional, creative, educational, technical, and research). Section E included a multi-select item on challenges experienced when using AI, which was used to measure alignment debt.

\subsection{Participants and Recruitment}

A non-probability convenience sampling strategy was used, with recruitment conducted through targeted social media advertising (X/Twitter and Facebook) and organic sharing in university, professional, and technology networks. Eligibility criteria required participants to be 18 years or older, currently residing in one of the study countries, and to have used an AI conversational tool.

The survey was initially deployed in Kenya, Nigeria, Ghana, and Senegal, yielding 2,327 responses. However, Ghana ($n=12$) and Senegal ($n=54$) were excluded from analysis due to insufficient sample sizes for meaningful prevalence or comparative inference. The final analytical sample therefore comprised 411 participants from Kenya ($n=282$) and Nigeria ($n=129$).

Alignment debt analyses used the measurable subsample ($n=385$). Twenty-six respondents (6.3\%) were excluded from debt scoring based on pre-registered rules: six who selected only the privacy item (“Shares my data without consent”), which is conceptually distinct from alignment debt, and twenty who selected only the free-text “Other” option. The latter group was excluded regardless of whether free-text responses could be post-hoc mapped to a category, to ensure consistent measurement conditions and avoid coder-dependent reclassification. Descriptive and verification context statistics use the full sample ($n=411$) where appropriate.

\subsection{Alignment Debt Measurement}

Alignment debt was measured using a multi-select item asking: “In general, what challenges and/or concerns do you have with using AI tools?” (Section E of the survey). Each selected challenge was mapped to one of the four debt categories defined in the alignment debt taxonomy (Table~\ref{tab:taxonomy_operationalisation}). A respondent was coded as experiencing a debt type if they selected at least one item within that category (binary OR logic). Cumulative alignment debt was calculated as the count of distinct debt categories present (range: 0-4), capturing the extent to which different forms of misalignment co-occur.

The option “Shares my data without consent” was treated as conceptually distinct from alignment debt, reflecting privacy and trust concerns rather than contextual misfit. Respondents who selected only this option (\textit{n}=6) were therefore excluded from alignment debt scoring. Respondents who selected only “Other, specify” (\textit{n}=20) were also excluded unless their text response could be reliably mapped to an existing category. After this exclusion and coding, 385 respondents (93.7\%) formed the analytical sample used in alignment debt prevalence, cross-country comparisons, and verification analyses.

\subsection{Data Quality}

Multiple quality control checks were implemented. IP addresses were tracked to identify and remove duplicate responses. For IP addresses appearing $\leq$3 times, one response was retained if it met duration and quality criteria (recognising that household members might legitimately use the same WiFi). For IP addresses appearing $>$3 times in rapid succession, all associated responses were flagged as potential fraud and excluded. Given the length of the survey, responses completed in under 6 minutes (median completion time: 8.5 minutes) were flagged for manual review. Open-ended responses were examined for gibberish or off-topic content. Participants confirmed current residence through a verification question. After applying these procedures, the final valid analytical sample comprised $N=411$ respondents.

\subsection{Participant Characteristics}

Table~\ref{tab:demographics} summarises the demographic and usage profile of the sample. The respondent pool skews young  (95.4\% under 35) and highly educated (84.9\% tertiary or postgraduate), consistent with current patterns of early AI adoption in Kenya and Nigeria. Most participants use AI tools daily and primarily access them on mobile devices. This demographic profile is consistent with early adopter populations and likely represents the most advantaged AI users in these contexts—those with highest digital literacy, best infrastructure access, and strongest English proficiency.  If alignment debt is prevalent even among this group, which is best positioned to benefit from AI, it is likely to be more severe in wider populations with lower bandwidth access, less English fluency, or less experience with AI systems.

\begin{table}[h]
\centering
\footnotesize
\caption{Sample Demographics}
\label{tab:demographics}

\small
\begin{tabular}{lccc}
\toprule
\textbf{Characteristic} & \textbf{Kenya} & \textbf{Nigeria} & \textbf{Overall} \\
 & (N=282)& (N=129)& (N=411) \\
\midrule
\textbf{Age} & & & \\
\quad 18-24 years & 61.0\%& 39.5\%& 54.3\%\\
\quad 25-34 years & 36.2\%& 51.9\%& 41.1\%\\
\quad 35-49 years & 2.5\%& 7.8\%& 4.1\%\\
\quad 50-64 years & 0.4\%& 0.8\%& 0.5\%\\
\midrule
\textbf{Education} & & & \\
\quad Secondary or less & 14.2\%& 17.1\%& 15.1\%\\
\quad Tertiary & 61.7\%& 51.9\%& 58.6\%\\
\quad Graduate/postgrad & 24.1\%& 31.0\%& 26.3\%\\
\midrule
\textbf{Top 3 Occupations} & & & \\
 \quad Student& 18.4\%& 20.2\%& 19.0\%\\
\quad Technology \& IT& 17.0\%& 10.9\%& 15.1\% \\
\quad Education& 11.7\%& 8.5\%& 10.7\%\\
\midrule
\textbf{AI Usage Frequency}& & & \\
\quad Everyday & 81.2\%& 75.2\%& 79.3\%\\
\quad Weekly & 16.7\%& 17.1\%& 16.8\%\\
\quad $\leq$2 weeks& 7.8\%& 2.1\%& 3.9\%\\
\midrule
\textbf{Most Used Tools} & & & \\
\quad ChatGPT & 90.1\%& 73.6\%& 84.9\%\\
\quad Meta AI & 57.0\%& 59.7\%& 57.9\%\\
\quad Google Gemini & 57.4\%& 27.9\%& 48.2\%\\
\midrule
\textbf{Paid Version Users} & 31.2\%& 17.8\%& 27.0\%\\
\bottomrule
\end{tabular}
\end{table}

\subsection{Data Quality}

Multiple quality control checks were implemented. IP addresses were tracked to identify and remove duplicate responses. For IP addresses appearing $\leq$3 times, one response was retained if it met duration and quality criteria (recognising that household members might legitimately use the same WiFi). For IP addresses appearing >3 times in rapid succession, all associated responses were flagged as potential fraud and excluded. Given the length of the survey, responses completed in under 6 minutes (median completion: 8.5 minutes) were flagged for manual review. Open-ended responses were examined for gibberish or off-topic content. Participants confirmed current residence through a verification question.

\subsection{Analytical Approach}

Analyses were structured around the three research questions and used the measurable subsample (\textit{N}=385) for all alignment debt analyses. Descriptive statistics that did not require debt classification (e.g., tool usage and verification behaviours) used the full sample (\textit{N}=411) and are clearly indicated when reported.

\subsubsection{RQ1: Prevalence of Alignment Debt}
For each debt type, we calculated the proportion of respondents who experienced at least one mapped challenge within that category (binary OR coding). Prevalence estimates are reported with 95\% Wilson score confidence intervals to provide accurate coverage with moderate sample sizes. Cumulative burden was operationalised as the count of debt types present (0-4), and its distribution is summarised using the mean, standard deviation, median, and the proportion of respondents at each burden level.

\subsubsection{RQ2: Cross-Country Variation}
To examine whether the prevalence of each debt type differed between Kenya and Nigeria, we used Pearson’s chi-square tests on 2×2 contingency tables. Fisher’s exact test was applied in cases where expected cell counts were below 5. Effect sizes are reported using Cramér's \(V\), with interpretation guided by both statistical significance and magnitude. Because the four comparisons address distinct constructs and risks of overcorrection outweigh interpretive benefit, we do not apply family-wise error correction in this block; however, exact \(p\)-values are reported transparently.

\subsubsection{RQ3: Debt and Verification Behaviour}
We conceptualise verification as a coping behaviour and analyse two outcomes:

\textit{Verification propensity} (binary): Respondents were classified as verifiers if they reported cross-checking AI outputs with any external source; those selecting only “I find the answers from AI tools satisfactory” were coded as non-verifiers. Three respondents with missing verification data were conservatively coded as non-verifiers. Chi-square tests assessed whether verification propensity differed by debt type, with Holm-Bonferroni correction across the four parallel tests. Cramér's \(V\) with 95\% CIs is reported, and we interpret differences of 5 percentage points or greater as practically meaningful.

\textit{Verification intensity} (continuous): We analysed the number of distinct verification sources used using Mann-Whitney \(U\) tests, reporting rank-biserial \(r\) with a practical difference threshold of 0.3 sources. A zero-inflation check confirmed that observed intensity patterns were not driven by non-verifiers alone.

To examine whether cumulative burden predicts verification behaviour, we computed Spearman’s \(\rho\) across the full 0-4 debt range. Because the four-type cell was small, a pre-registered sensitivity analysis excluded that group and re-tested monotonic trend using both Spearman’s \(\rho\) and the Cochran-Armitage test; both confirmed the positive relationship.

\section{Results}

We address our research questions by examining how often each debt type occurs (RQ1), how patterns vary across the two contexts (RQ2), and how users respond behaviourally through verification (RQ3). All alignment-debt analyses use the measurable subsample (\textit{N}=385); descriptive verification statistics use the full sample (\textit{N}=411) where noted.

\subsection{Prevalence of Alignment Debt (RQ1)}

Alignment debt is widespread. Among respondents measurable on the four-category taxonomy (\textit{N}=385), every individual experienced at least one form of misalignment. Cultural and Linguistic debt is the most common, affecting 51.9\% (95\% CI: 47.0-56.9, \textit{n}=200). These users report accent and dialect misunderstandings (29.0\%), lack of cultural awareness in responses (21.2\%), or perceived judgement or bias in tone (8.5\%). These cases reflect situations where the system's language or cultural assumptions do not align with those of the user, leading to conversational friction or responses that feel irrelevant, incorrect, or out of place.

Infrastructural debt affects 43.1\% of users (\textit{n}=166, 95\% CI: 38.3-48.1). This debt reflects cases where AI tools assume stable, high-capacity connectivity that does not match users' reality. Users report slow response times (16.3\%) and requirements for stronger or more reliable internet signals (29.2\%), indicating that access to AI is constrained by bandwidth and network stability.

Epistemic debt affects 33.8\% of users (\textit{n}=130, 95\% CI: 29.2-38.6). This category reflects cases where users encounter outputs they perceive as incorrect, misleading, or insufficiently supported. Within this group, 19.0\% report the absence of source explanations, 9.2\% report wrong answers, and 8.8\% report misinformation. These patterns indicate that, for many users, AI outputs often cannot be taken at face value and may require additional checking .

Interaction debt is least common at 14.0\% (\textit{n}=54, 95\% CI: 10.9-17.9). This reflects cases where users report difficulty formulating prompts or being understood by the system. While less prevalent than other debt types, it still represents a meaningful form of additional work for affected users, who must adjust how they express requests to obtain useful outputs.

\begin{figure}[t]
  \centering
  \includegraphics[width=0.70\linewidth]{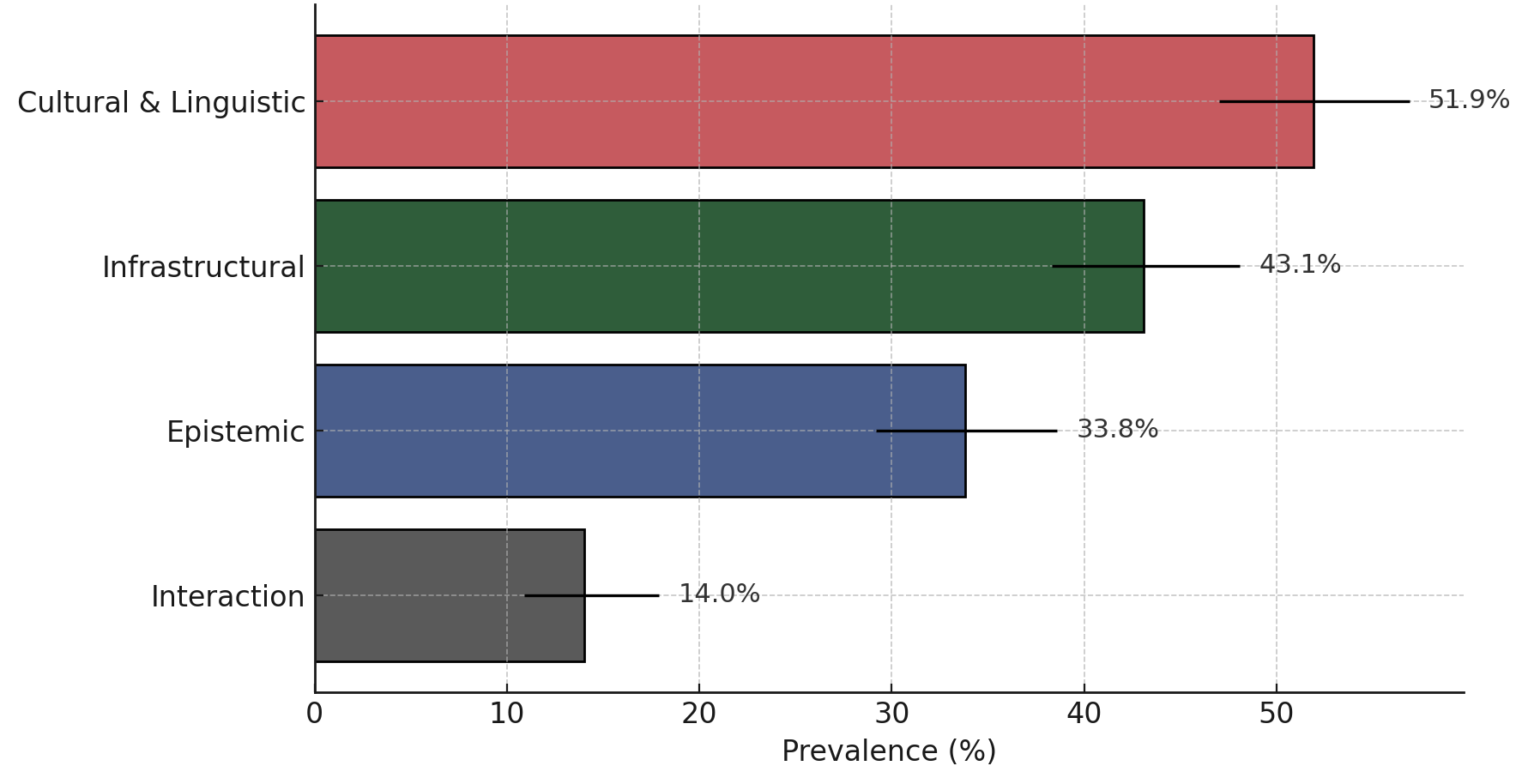}
  \caption{Alignment Debt Prevalence (\textit{N}=385) \\
  Across the full sample (\textit{N}=411), 97.6\% report at least one challenge, indicating that misalignment is a near-universal condition of use.}
  \label{fig:prevalence}
\end{figure}

\subsection{Cross-Country Variation (RQ2)}

Patterns of alignment debt vary across Kenya and Nigeria, but not uniformly across debt types. Infrastructural debt is significantly more common in Kenya (47.0\%) than in Nigeria (33.9\%) (\(\chi^2=5.14\), \textit{p}=0.023, Cramér's V=0.116). Connectivity appears to shape this difference: 33.3\% of Kenyan users report weak signal compared to 26.1\% of Nigerians, and slow loading affects 20.4\% of Kenyans versus 10.4\% of Nigerians.

Interaction debt diverges more sharply: 17.4\% in Kenya versus 6.1\% in Nigeria (\(\chi^2 = 7.66\), \textit{p} = 0.006, Cramér's V = 0.141). This 11.3 percentage point difference is the largest cross-country divergence observed. Kenya’s sample skews younger (61\% aged 18-24) and has a somewhat higher proportion of technology professionals (17.0\% vs 10.9\%), whereas Nigeria’s sample includes more postgraduate respondents (31.0\% vs 24.1\%). While we cannot infer causality from this data, these compositional differences provide context for interpreting variation in Interaction Debt across the two countries.

In contrast, Cultural/Linguistic and Epistemic debts are broadly similar between Kenya and Nigeria. This indicates that these forms of misalignment may be shaped by factors that are shared across the two contexts rather than country-specific infrastructure differences. However, our data do not speak to the underlying causes of these similarities, and further work is required to examine whether they reflect model training data, interface design, or broader representational patterns.

\begin{figure}[t]
  \centering
  \includegraphics[width=0.70\linewidth]{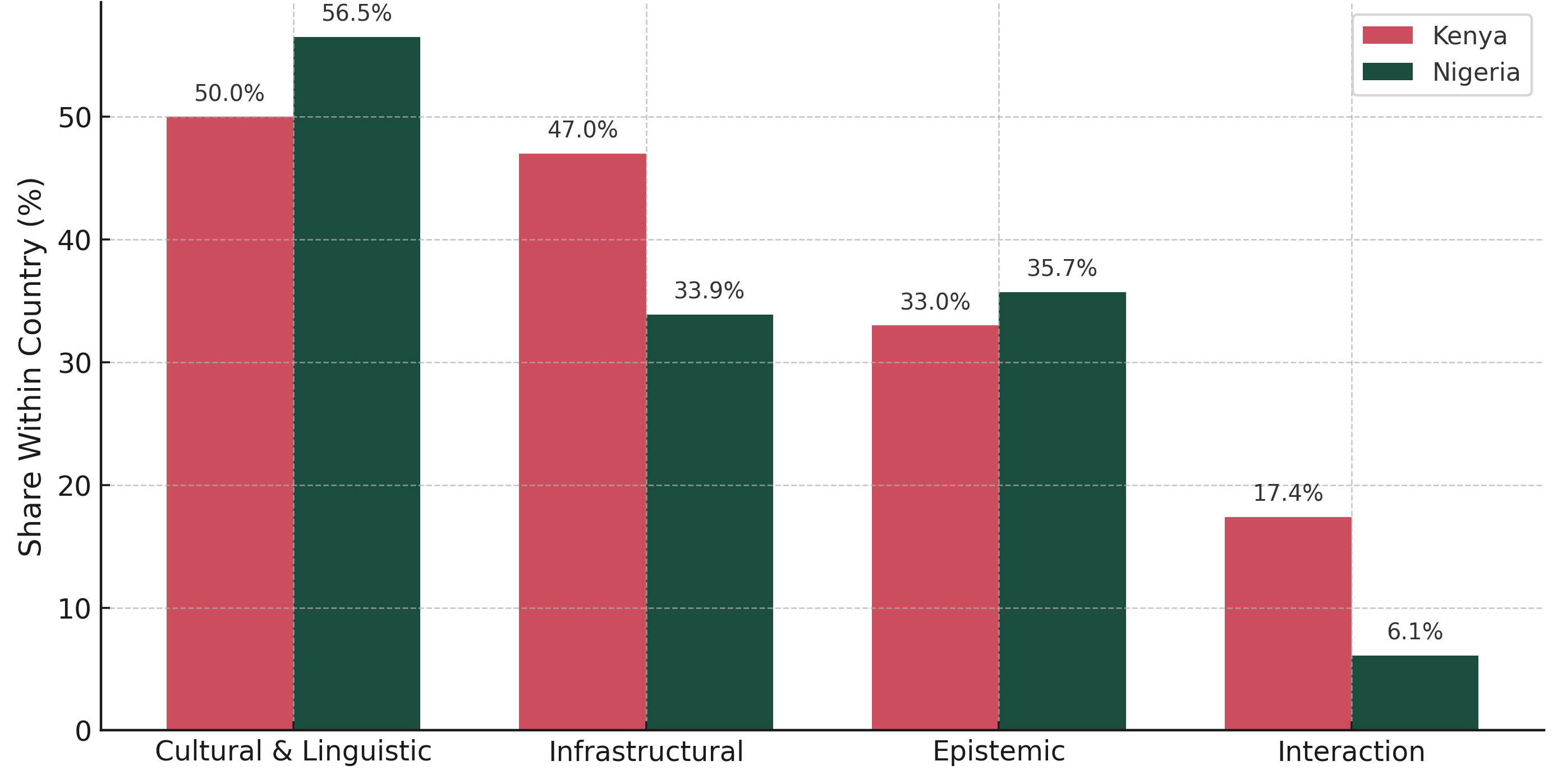}
  \caption{Alignment Debt Prevalence by Country (Kenya $n{=}270$ vs. Nigeria $n{=}115$ ) \\
  Infrastructural: $\chi^2{=}5.14$, $p{=}0.023$, Cramér's $V{=}0.116$; Interaction: $\chi^2{=}7.66$, $p{=}0.006$, Cramér's $V{=}0.141$.}
  \label{fig:country}
\end{figure}

\subsection{Cumulative Debt}

Although most users experience a single debt type, 34.0\% experience multiple. Sixty-six percent experience one type, 26.8\% experience two, 5.7\% experience three, and 1.6\% experience all four. Mean cumulative burden is 1.43 types per user (SD 0.67).

\subsection{Verification Behaviour}

Verification is widespread. 84.6\% (\textit{n}=348) cross-check AI outputs using external sources. Only a small minority (15.3\%) accept AI outputs without external checks. For the vast majority who do verify, Google serves as the primary tool (64.5\%), followed by academic or authoritative databases (24.1\%) and Wikipedia (13.9\%). 

\subsection{Debt Type and Verification Propensity (RQ3)}

Epistemic debt is strongly associated with verification propensity. Users experiencing epistemic debt verify at 91.5\%, compared to 80.8\% among those without (\(\chi^2 = 6.77\), \textit{p} = 0.037 corrected, Cramér's V = 0.133).

Other debt types did not demonstrate meaningful associations with verification. Cultural and Linguistic debt showed only a small, non-significant difference in verification rates (86.0\% with debt vs 82.7\% without; \(\chi^2 = 0.56, p = 0.906\)). Interaction debt similarly showed no association (83.3\% with vs 84.6\% without; \(\chi^2 = 0.00, p = 0.973\)). These non-effects suggest that verification is not a universal coping response to misalignment, but one that is mobilised specifically when users question the reliability of system outputs.

Infrastructural debt shows a small and non-significant negative association with verification (80.7\% with debt vs 87.2\% without; \(\chi^2 = 2.55, p = 0.331\), Cramér's V = 0.082). The direction of this pattern is noteworthy: users facing connectivity constraints verify slightly less, rather than more, likely because verification itself requires additional network access. This suggests that some burdens cannot be compensated for through verification, since the conditions required to verify are the same conditions that are already strained.

\subsection{Cumulative Debt and Verification Intensity}

Beyond whether users verify, cumulative debt predicts how intensively they verify. Users with one debt type consult an average of 1.48 sources when verifying, rising to 1.64 sources for two types, 1.95 for three types, and 3.50 for all four. Users experiencing all four debt types consult more than double the baseline. Spearman's rank correlation confirms this monotonic trend (Spearman’s \(\rho\) = 0.147, \textit{p} = 0.004).

This dose-response relationship demonstrates compounding burden. Users facing multiple debts must check more extensively to achieve confidence in information. The median remains 1.0 source for users with one through three debts, indicating most users check exactly one source. However, the mean increases because a subset checks multiple sources intensively.

Verification patterns do not differ meaningfully by country, suggesting that once misalignment is perceived, compensatory work looks similar across contexts.

\begin{table}[h]
\centering
\footnotesize
\caption{Verification Intensity by Cumulative Debt (\textit{N} = 385)}
\label{tab:intensity}
\begin{tabular}{@{}lrrrr@{}}
\toprule
\textbf{Debt} & \textbf{\textit{N}} & \textbf{Verify} & \textbf{Mean} & \textbf{Median} \\
\textbf{Count} & & \textbf{\%} & \textbf{Sources} & \textbf{Sources} \\
\midrule
1 debt & 254 & 82.3 & 1.48 & 1.0 \\
2 debts & 103 & 88.3 & 1.64 & 1.0 \\
3 debts & 22 & 95.5 & 1.95 & 1.0 \\
4 debts & 6 & 66.7 & 3.50 & 4.0 \\
\bottomrule
\multicolumn{5}{l}{\footnotesize Spearman \(\rho\) = 0.147, \textit{p} = .004} \\
\end{tabular}
\end{table}
\begin{figure}[t]
  \centering
  \includegraphics[width=0.70\linewidth]{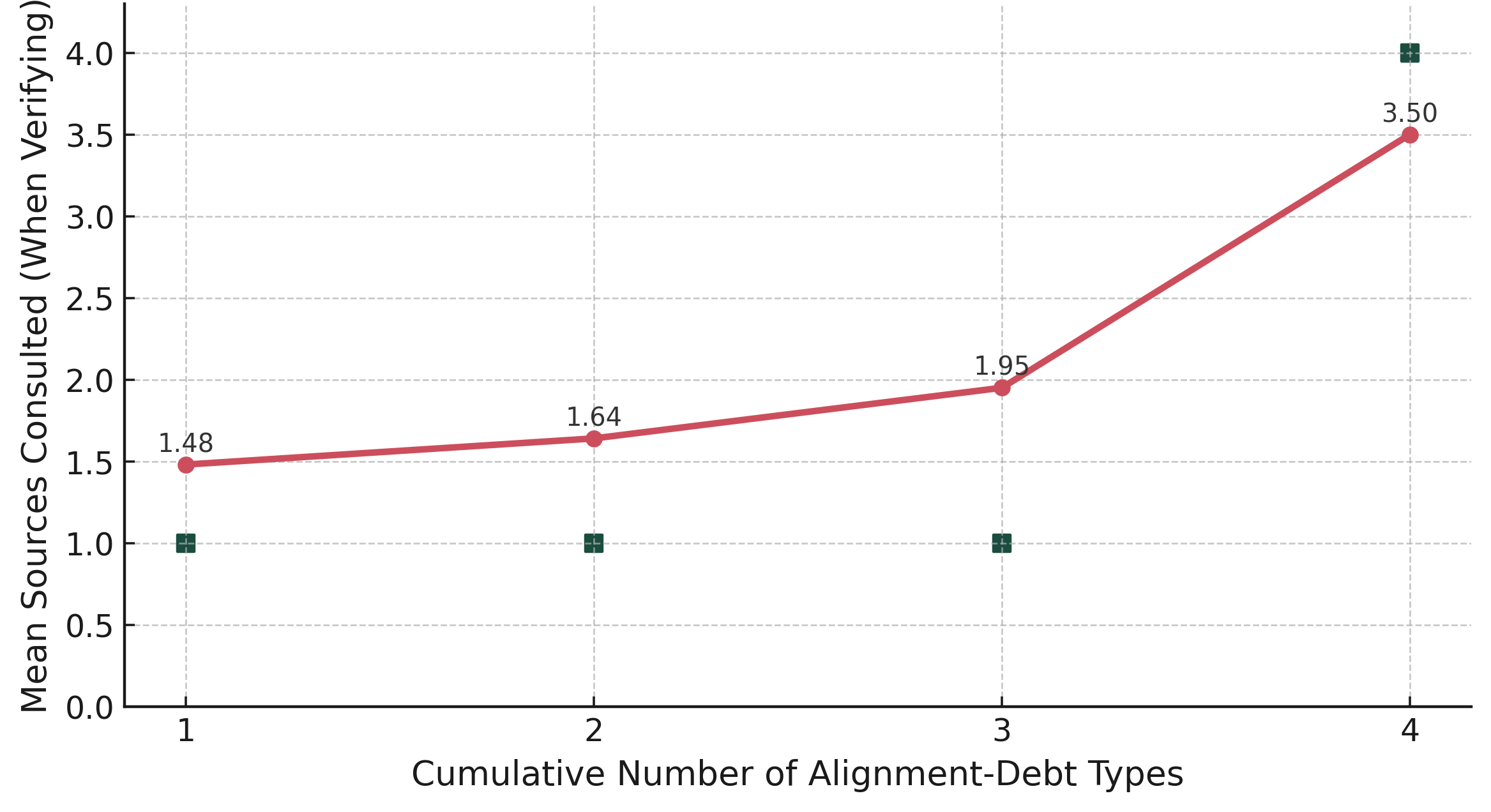}
  \caption{Verification Intensity by Cumulative Debt.\\
  Circles show means, squares show medians }
  \label{fig:verify-intensity}
\end{figure}

Taken together, these findings suggest that verification is triggered primarily by epistemic uncertainty, but once verification is initiated, the intensity of that verification increases with the overall burden of misalignment a user experiences.

\section{Discussion}

The findings indicate that alignment debt is not a sporadic inconvenience but a routine condition of everyday AI use in Kenya and Nigeria. They also challenge the assumption that widespread adoption implies successful alignment. Users in our sample adopt and actively use AI tools not because these systems are well attuned to their contexts, but because they provide value despite misalignment. Adoption, therefore, should not be interpreted as frictionless use, as it often coexists with substantial compensatory labour.

Rather than evaluating systems only in terms of aggregate accuracy or fairness metrics, our analysis foregrounds the labour users perform to make globally deployed AI systems workable in their local contexts. We interpret this in four steps: first, by examining alignment debt as a form of user labour; second, by situating this labour in structural conditions of representation, infrastructure, and interaction; third, by outlining design implications that reduce burden without shifting responsibility back to users; and finally, by considering how alignment debt can inform emerging African AI governance frameworks.

\subsection{Alignment Debt as User Labour}

Every respondent in our measurable sample experienced at least one form of alignment debt, and many experienced multiple forms. Cultural and Linguistic and Epistemic debts were the most common. These patterns indicate that users cannot reliably take system outputs at face value. Instead, they often need to rephrase requests, interpret meaning through contextual knowledge, or seek external confirmation before acting. In Kenya and Nigeria, multilingual communication and routine code-switching are part of everyday expression across English, Swahili, Pidgin, Yoruba, and other local languages. Models trained primarily on standardised Western English implicitly constrain this linguistic repertoire. When accent or dialect is misrecognised, users are not simply correcting a lexical error; they are adjusting how they are allowed to sound. The labour here is therefore not only cognitive but also expressive, involving subtle forms of linguistic self-modification.

This labour is not a function of users' lack of ability or familiarity with AI systems. Rather, it reflects the distance between model assumptions and the contexts in which they are deployed. The cultural frames, linguistic repertoires, and infrastructural conditions that structure everyday communication in Kenya and Nigeria are not embedded in the data that shape current models. As a result, the burden of alignment falls on users, who must bridge that gap through additional effort. Classic work on user burden conceptualises such effort as cumulative and a predictor of abandonment \cite{Suh2016}. Our contribution is to show how this burden manifests specifically in generative AI use.

Two empirical patterns support this interpretation. First, verification is mobilised selectively, and most strongly, when respondents encounter Epistemic debt. This is consistent with existing evidence that trust calibration is resource intensive and triggered when reliability is uncertain \cite{Wischnewski2023}. Second, verification intensity scales with cumulative debt: users experiencing multiple forms of debt consult more sources to reach a point where acting on the output feels safe. This dose-response pattern indicates that alignment debt is an accumulation of compensatory labour that increases as misalignment deepens.

Finally, users experiencing Infrastructural debt verify \emph{less}, not more. Verification itself requires bandwidth, stable connectivity, time and access to additional sources, which are the same resources already constrained when infrastructure is unreliable. This shows that the capacity to compensate for misalignment is not evenly distributed, as some users are structurally prevented from performing the labour that alignment debt would otherwise require. 

\subsection{Why These Burdens Persist}

The distribution of burdens aligns with known structural factors in how AI systems are built. Cultural and Linguistic debt persists across both countries, despite demographic differences, reflecting that the systems they interact with lack representations of their languages, dialects, and cultural references. This supports the body of evidence that African linguistic and cultural contexts remain under-represented in the corpora that shape contemporary large models \cite{Nekoto2020,Graham2014}.

The pattern observed for infrastructural constraints further highlights that burden is shaped by structural conditions.  Systems assume stable, low-latency connectivity and low marginal cost of access; conditions that do not hold for many users in local African contexts. In these cases, the very conditions that create misalignment also limit the user’s capacity to compensate for it. The burden is therefore not only cognitive; it is materially patterned.

The layering of debts across language, context, reliability, and infrastructure suggests that misalignment is cumulative. It is not one challenge but a patterned set of frictions arising from where models are built, the data they learn from, and the use environments into which they are deployed.

\subsection{Design Implications}

Our findings point to a pragmatic design agenda that reduces the need for users to perform compensatory labour and shifts the responsibility for alignment upstream to designers and platforms. That said, it is important to note that this is not an engineering checklist but a set of trade-offs that will look different depending on task, sector, and local conditions.  

For cultural and linguistic debt, we suggest two moves. First, invest in representation by building participatory pipelines and community-led data projects to enable models integrate the linguistic varieties present in Kenya, Nigeria, and beyond.  In practice, this might mean supporting initiatives that document code switching and dialectal forms, funding corpora development, and licensing local news and scholarly sources for retrieval.  Second, improve the interface of AI systems to acknowledge uncertainty when dialect or phrasing is ambiguous, and request clarification instead of flattening meaning or silently defaulting to Western interpretations.  These steps both reduce the need for users to change their speech patterns and create a clearer feedback loop for model improvement.

To reduce infrastructural debt, rather than requiring users to buy faster plans, systems should be designed to work better with limited bandwidth and intermittent connectivity. Design choices such as low bandwidth modes, progressive loading, explicit pre-query data estimates so users can choose whether to proceed, queueing requests when a device is offline, and offering compact models that can run locally on device can reduce access inequities. Although these design choices are technically achievable, implementation is highly dependent on business incentive. Companies typically optimise for their biggest or most profitable user markets, which tend to be wealthy countries with cheap data and stable connectivity. If a feature primarily benefits users in places where revenue per user is low and data costs are high, the company has little financial motivation to prioritise it. So, without deliberate commitment from these companies (or external pressure such as policy or public procurement requirements), there is no market incentive pushing companies to design for low-bandwidth African or Global South contexts. This means that the places where these features are most needed are the places least likely to receive them.

The cost of verification associated with epistemic debt can be lowered if systems make uncertainty levels clear rather than implied. This could be done by attaching inline sources to factual claims so users can open the precise documents supporting a statement or by showing confidence bands and flagging low-confidence responses that rely on sparse data. Complementing those interface cues, systems should be configured to prefer local and regional sources when queries mention African geographic or cultural terms.  This will not solve all errors, and it may sometimes increase false reassurance if the local sources are weak themselves; still, prioritising regional material is a concrete step that could reduce the burden of external verification for many users or even help them judge when verification is still necessary.

Prompt engineering should not be framed as a user skill deficit in a bid to tackle interaction debt, as this shifts responsibility to the users. Instead, systems can integrate these simple design choices to lower the required skill threshold prompt writing. Provide context-aware templates  aligned with common local communication styles; ask short, interactive clarifying questions after an initial response so users do not have to predict every detail up front; and provide short, localised examples of good prompts for common tasks.  These supports should be optional not prescriptive, so as to avoid constraining how users choose to express themselves.

Across all these design domains, measurement is key. Designers and evaluators should track metrics such as verification rates, verification time, and data cost per task. These operational metrics can be used as product KPIs so that teams optimise for lower user burden as well as higher system performance.  Finally, none of these measures are neutral. Most interventions have distributional effects. Offline model variants, for example, can reduce data costs but may lag in updates. We therefore recommend iterative, participatory evaluation with affected communities so improvements actually reduce user burden effectively.

\subsection{Policy and Governance}

Without complementary policy action, design improvements may struggle to scale or not happen at all. Policy can nudge platforms to report and mitigate user burden through different routes. One sensible route is to require alignment checks as part of procurement and sectoral approvals. When a government buys an AI service for education or health, the vendor could be asked to report indicators that capture expected user burden in the intended population, including typical verification rates and time, accent and dialect handling, connectivity sensitivity, and expected data consumption for common tasks. Framing contextual alignment as a routine procurement requirement makes it something that is planned for, not assumed or hoped for. It also creates a direct incentive for developers to develop low bandwidth or offline alternatives.

Public infrastructure investment plays a second, enabling role. Governments and multilateral agencies can seed the public data and compute resources that companies currently centralise; public funding for African language and domain specific corpora, national knowledge graphs, and shared compute would lower the barrier to developing locally adapted models. None of this removes the need for private sector innovation, but it redistributes who has the capacity to shape contextual alignment.

Third, standards bodies should extend conformance frameworks to include user burden metrics. This does not require turning every performance indicator into a legal mandate; alternatively, it involves embedding a requirement that fairness claims be accompanied by evidence about user burden. Incorporating impact assessments that require vendors to report user burden estimates and mitigation plans alongside accuracy and bias tests would achieve this.

At the regional level, the African Union’s \textit{Continental AI Strategy} prefigures these policy recommendations by emphasising equitable access, local capacity building, and the development of contextually relevant data and evaluation practices \cite{AfricanUnion2024}. Although it does not specify how contextual alignment should be measured practically, it makes a clear point that AI systems deployed in Africa should reflect local social, cultural, and infrastructural realities. National strategies appear to be evolving in a similar direction. Kenya's National AI Strategy underscores public sector capacity building, localisation, shared data infrastructure, and governance frameworks as central to responsible deployment \cite{KenyaAI2025}. Nigeria’s National AI Strategy places more emphasis on strengthening domestic innovation ecosystems and guiding responsible use in state services \cite{NigeriaAI2025}. 

These strategies are early stage and more aspirational than operational. Hence, they outline ambitions but do not resolve questions of \textit{how} alignment should be measured, \textit{who} should be held accountable when it fails, and \textit{what} enforcement might look like in diverse national contexts. Nevertheless, their very existence shows a recognition of contextual alignment as not only a design concern, but governance one also.

As these strategies mature, they could provide institutional footholds for experimentation. Public agencies could, for instance, test alignment debt indicators in limited pilots before wider adoption. This would help clarify whether the notion of user burden can be actionable in bureaucratic practice, or whether it will remain a moral aspiration that policy documents gesture toward but seldom implement. The task ahead is not to articulate new visions of responsible AI, but to translate these existing commitments into tangible governance practices that measure and redistribute the effort of making AI usable in African contexts.

\subsection{Limitations}

It is requisite to take the limitations of this study into account while interpreting these findings. Our sample skews heavily toward younger, well educated participants; 95\% under 35 and 86\% with tertiary or higher levels of education. This likely means the burdens reported here are lower than what might be observed among older users or those with less formal education, who often face steeper learning curves and may have lower tolerance for misalignment. As participation required internet access and prior familiarity with AI tools, those most marginalised by unstable connectivity or linguistic exclusion were, almost by design, least represented. The patterns captured here should therefore be viewed as describing the upper bound of adaptation rather than the full spectrum of the African AI experience.

Measures in this study are self-reported and capture perception rather than instrumented behaviour. Ergo, recall and social desirability biases such as participants overstating careful behaviours like verification or understating less socially desirable behaviours like prompting difficulty, can be introduced \cite{VanDeMortel2008}. Although perception is central to understanding lived burden, future studies combining self-reports with behavioural or telemetry data could help disentangle what users think they do from what interaction logs show. Without behavioural traces or task-linked telemetry, we cannot triangulate perceptions against observed interaction patterns or time spent on task.

Considering geography, the study covers only two anglophone countries, Kenya and Nigeria, which differ from one another and from the rest of the continent’s linguistic and infrastructural diversity. We cannot generalise Africa’s linguistic and infrastructural diversity from a single survey. What appears as “common African experience” here should be read instead as a situated baseline shaped by anglophone digital infrastructures and particular histories of platform adoption.

Despite these limits, the findings establish a credible empirical base for theory building around user-side burden and offer a starting point for more comparative and time sensitive studies that trace how alignment debt evolves with AI systems.

\subsection{Future Directions}

Treating alignment debt as measurable opens several paths for research experimentation. Below we outline a set of complementary studies that would move the field from description to measurement, and measurement to institutional and economic accountability. 

\paragraph{Temporal Measurement.}
Understanding alignment debt on a deeper level will require longitudinal studies that follow users for several months. This could help trace how coping strategies evolve: does the burden fade as familiarity grows, or does burden accumulate until abandonment? Results could clarify whether alignment debt is a problem of system maturity which resolves as models improve or a structural feature of deployment without local grounding. It would also show whether interventions to reduce burden need to be permanent fixtures or adaptive supports that can be phased as capacity and representation improve. 

\paragraph{Instrument Validation and Operationalisation.}
Future work should focus on building and validating short form instruments that measure alignment debt reliably and with minimal burden on participants. One useful next step could involve cross-validating self-reports with behavioural traces like keystroke data, task duration, or bandwidth usage, to calibrate accuracy without invading users' privacy. These instruments could then be embedded in public procurement pilots or platform telemetry, helping translate the concept from research to practice. Extending measurement to multimodal systems, such as speech and vision in healthcare or education could also uncover new facets of alignment debt. These instruments should be co-designed with local researchers and user groups within affected communities to ensure they reflect experiences of burden which are locally salient. 

\paragraph{Expanded Geography.}
Replicating this research in francophone West Africa, lusophone Southern Africa, and Arabic speaking North Africa would reveal whether alignment debt follows predictable patterns or emerges differently depending on language and infrastructure. Do societies with stronger oral traditions experience debt differently? How do countries where governments have invested heavily in digital infrastructure measure against those with little or no such infrastructural investment? From these comparisons we may find that some forms of burden are universal while others are highly contingent. That distinction would help sharpen what contextual alignment actually means and where to focus limited resources. Further exploration could also include stratifying by rural-urban location, connectivity, gender, disability, and data pricing exposure.

\paragraph{Developer and Industry Perspectives.}
Understanding how companies that build AI systems perceive alignment debt may be as important as measuring it among those who use these systems. Interviews and short ethnographies with engineers, product managers, localisation teams, trust and safety reviewers, and policy staff could clarify how Global South usability challenges are framed, prioritised, and resourced internally; whether as edge cases or market expansion issues. These insights would reveal the thresholds of evidence that trigger action and the incentive structures that reward certain forms of improvement while ignoring others. Research here should not only diagnose organisational inertia but experiment with levers that could shift it: procurement rules, transparency incentives, or burden reporting tied to performance reviews. Such work would bridge the gap between knowledge and adoption, identifying which accountability mechanisms can make user burden visible to companies accustomed to optimising for other metrics.

Collectively, these directions suggest a research programme for AI systems that examine alignment as an evolving relationship between design, infrastructure, and society. The goal is to build the empirical and institutional tools to make that relationship measurable, and, ultimately, negotiable.

\subsection{Conclusion}

The central implication of this study is that fairness in AI cannot be judged by model metrics alone. It must also be judged by the burden users are saddled with and labour they perform in an attempt to make AI systems usable in their contexts. Alignment debt gives us a lens to understand this labour; to name the hidden effort involved in bridging the gap between design assumptions and reality.

The question now is not about whether alignment debt exists but whether stakeholders (companies, regulators, researchers, and communities) will take action to address it. Addressing it will entail deliberate investment in local data infrastructures, design choices that lower user burden and effort, and governance frameworks that treat user burden as a core metric of responsible AI. AI systems will only begin to meaningfully align with the Global South contexts in which they are used when alignment itself is treated as a collective obligation.

\section*{Author Contributions}

\textbf{Cumi Oyemike (C.O.)} conceived the alignment debt concept and framework, co-developed the study design, conducted advanced data cleaning and all statistical analyses, and wrote the manuscript.\\
\textbf{Elizabeth Akpan (E.A.)} led the conceptual design of the study and the design of the survey instrument and research protocol.\\
\textbf{Pierre Hervé-Berdys (P.H.-B.)} led participant recruitment, coordinated data collection, and performed initial data cleaning.

\section*{Competing Interests}

The authors declare no competing interests.

\section*{Correspondence}

Correspondence and requests for materials should be addressed to \textbf{Cumi Oyemike}.

\bibliographystyle{unsrtnat}
\bibliography{references}

\begin{thebibliography}{59}
\providecommand{\natexlab}[1]{#1}
\providecommand{\url}[1]{\texttt{#1}}
\expandafter\ifx\csname urlstyle\endcsname\relax
  \providecommand{\doi}[1]{doi: #1}\else
  \providecommand{\doi}{doi: \begingroup \urlstyle{rm}\Url}\fi

\bibitem[Suh et~al.(2016)Suh, Shahriaree, Hekler, et~al.]{Suh2016}
Hyewon Suh, Nina Shahriaree, Eric~B. Hekler, et~al.
\newblock Developing and validating the user burden scale: A tool for assessing user burden in computing systems.
\newblock In \emph{Proceedings of the 2016 CHI Conference on Human Factors in Computing Systems}, pages 3988--3999, 2016.
\newblock \doi{10.1145/2858036.2858448}.

\bibitem[Buolamwini and Gebru(2018)]{Buolamwini2018}
Joy Buolamwini and Timnit Gebru.
\newblock Gender shades: Intersectional accuracy disparities in commercial gender classification.
\newblock In \emph{Proceedings of the 1st Conference on Fairness, Accountability and Transparency}, volume~81 of \emph{Proceedings of Machine Learning Research}, pages 77--91, 2018.
\newblock URL \url{https://proceedings.mlr.press/v81/buolamwini18a.html}.

\bibitem[Obermeyer et~al.(2019)Obermeyer, Powers, Vogeli, et~al.]{Obermeyer2019}
Ziad Obermeyer, Brian Powers, Christine Vogeli, et~al.
\newblock Dissecting racial bias in an algorithm used to manage the health of populations.
\newblock \emph{Science}, 366\penalty0 (6464):\penalty0 447--453, 2019.
\newblock \doi{10.1126/science.aax2342}.

\bibitem[Noble(2018)]{Noble2018}
Safiya~Umoja Noble.
\newblock \emph{Algorithms of Oppression: How Search Engines Reinforce Racism}.
\newblock NYU Press, 2018.

\bibitem[Bender et~al.(2021)Bender, Gebru, McMillan-Major, et~al.]{Bender2021}
Emily~M. Bender, Timnit Gebru, Angelina McMillan-Major, et~al.
\newblock On the dangers of stochastic parrots: Can language models be too big?
\newblock In \emph{Proceedings of the 2021 ACM Conference on Fairness, Accountability, and Transparency}, FAccT 2021, pages 610--623, 2021.
\newblock \doi{10.1145/3442188.3445922}.

\bibitem[Bolukbasi et~al.(2016)Bolukbasi, Chang, Zou, et~al.]{Bolukbasi2016}
Tolga Bolukbasi, Kai-Wei Chang, James Zou, et~al.
\newblock Man is to computer programmer as woman is to homemaker? debiasing word embeddings.
\newblock In \emph{Proceedings of the 30th International Conference on Neural Information Processing Systems}, NeurIPS 2016, page 4356–4364, 2016.
\newblock \doi{10.5555/3157382.3157584}.

\bibitem[Birhane(2020)]{Birhane2020}
Abeba Birhane.
\newblock Algorithmic colonization of africa.
\newblock \emph{SCRIPTed: A Journal of Law, Technology \& Society}, 17\penalty0 (2):\penalty0 389--409, 2020.
\newblock \doi{10.2966/scrip.170220.389}.

\bibitem[Mohamed et~al.(2020)Mohamed, Png, and Isaac]{Mohamed2020}
Shakir Mohamed, Marie-Therese Png, and William Isaac.
\newblock Decolonial {AI}: Decolonial theory as sociotechnical foresight in artificial intelligence.
\newblock \emph{Philosophy \& Technology}, 33\penalty0 (4):\penalty0 659--684, 2020.
\newblock \doi{10.1007/s13347-020-00405-8}.

\bibitem[Nekoto et~al.(2020)Nekoto, Marivate, Matsila, et~al.]{Nekoto2020}
Wilhelmina Nekoto, Vukosi Marivate, Tshinondiwa Matsila, et~al.
\newblock Participatory research for low-resourced machine translation: A case study in african languages.
\newblock In \emph{Findings of the Association for Computational Linguistics: EMNLP 2020"}, pages 2144--2160, 2020.
\newblock \doi{10.18653/v1/2020.findings-emnlp.195}.

\bibitem[Orife et~al.(2020)Orife, Kreutzer, Sibanda, et~al.]{Orife2020}
Iroro Orife, Julia Kreutzer, Blessing Sibanda, et~al.
\newblock Masakhane -- machine translation for africa.
\newblock In \emph{AfricaNLP Workshop, Proceedings of the International Conference on Learning Representations (ICLR)}, 2020.
\newblock \doi{10.48550/arXiv.2003.11529}.

\bibitem[Koenecke et~al.(2020)Koenecke, Nam, Lake, et~al.]{Koenecke2020}
Allison Koenecke, Andrew Nam, Emily Lake, et~al.
\newblock Racial disparities in automated speech recognition.
\newblock \emph{Proceedings of the National Academy of Sciences}, 117\penalty0 (14):\penalty0 7684--7689, 2020.
\newblock \doi{10.1073/pnas.1915768117}.

\bibitem[Winata et~al.(2023)Winata, Aji, Yong, et~al.]{Winata2023}
Genta Winata, Alham~Fikri Aji, Zheng~Xin Yong, et~al.
\newblock The decades progress on code-switching research in {NLP}.
\newblock In \emph{Findings of the Association for Computational Linguistics: ACL 2023}, pages 2936--2978, 2023.
\newblock \doi{10.18653/v1/2023.findings-acl.185}.

\bibitem[Olatunji et~al.(2023)Olatunji, Afonja, Yadavalli, et~al.]{Olatunji2023}
Tobi Olatunji, Tejumade Afonja, Aditya Yadavalli, et~al.
\newblock {AfriSpeech-200}: Pan-african accented speech dataset for clinical and general domain {ASR}.
\newblock \emph{Transactions of the Association for Computational Linguistics}, 11:\penalty0 1669--1685, 2023.
\newblock \doi{10.1162/tacl_a_00627}.

\bibitem[Sheng et~al.(2019)Sheng, Chang, Natarajan, et~al.]{Sheng2019}
Emily Sheng, Kai-Wei Chang, Premkumar Natarajan, et~al.
\newblock The woman worked as a babysitter: On biases in language generation.
\newblock In \emph{Proceedings of the 2019 Conference on Empirical Methods in Natural Language Processing and the 9th International Joint Conference on Natural Language Processing}, pages 3407--3412, 2019.
\newblock \doi{10.18653/v1/D19-1339}.

\bibitem[{African Union}(2024)]{AfricanUnion2024}
{African Union}.
\newblock African union continental {AI} strategy, 2024.
\newblock URL \url{https://au.int/en/documents/20240809/continental-artificial-intelligence-strategy}.

\bibitem[Asiedu et~al.(2024)Asiedu, Dieng, Haykel, et~al.]{Asiedu2024}
Mercy~Nyamewaa Asiedu, Awa Dieng, Iskandar Haykel, et~al.
\newblock The case for globalizing fairness: A mixed methods study on colonialism, {AI}, and health in africa.
\newblock In \emph{Proceedings of the 4th ACM Conference on Equity and Access in Algorithms, Mechanisms, and Optimization}, EAAMO '24. Association for Computing Machinery, 2024.
\newblock \doi{10.1145/3689904.3694708}.

\bibitem[Pasipamire and Muroyiwa(2024)]{Pasipamire2024}
Notice Pasipamire and Abton Muroyiwa.
\newblock Navigating algorithm bias in {AI}: Ensuring fairness and trust in africa.
\newblock \emph{Frontiers in Research Metrics and Analytics}, 9, 2024.
\newblock \doi{10.3389/frma.2024.1486600}.

\bibitem[Mahamadou et~al.(2024)Mahamadou, Ochasi, and Altman]{Mahamadou2024}
Abdoul Jalil~Djiberou Mahamadou, Aloysius Ochasi, and Russ~B. Altman.
\newblock Data ethics in the era of healthcare artificial intelligence in africa: An {U}buntu philosophy perspective.
\newblock \emph{arXiv}, 2024.
\newblock \doi{10.48550/arXiv.2406.10121}.

\bibitem[Wischnewski et~al.(2023)Wischnewski, Kr\"{a}mer, and M\"{u}ller]{Wischnewski2023}
Magdalena Wischnewski, Nicole Kr\"{a}mer, and Emmanuel M\"{u}ller.
\newblock Measuring and understanding trust calibrations for automated systems: A survey of the state-of-the-art.
\newblock In \emph{Proceedings of the 2023 CHI Conference on Human Factors in Computing Systems}, CHI '23, pages 1--24, 2023.
\newblock \doi{10.1145/3544548.3581372}.

\bibitem[Simkute et~al.(2025)Simkute, Tankelevitch, Kewenig, et~al.]{Simkute2025}
Auste Simkute, Lev Tankelevitch, Viktor Kewenig, et~al.
\newblock Ironies of generative {AI}: Understanding and mitigating productivity loss in human-{AI} interactions.
\newblock \emph{International Journal of Human–Computer Interaction}, 41\penalty0 (5):\penalty0 2898--2919, 2025.
\newblock \doi{10.1080/10447318.2024.2405782}.

\bibitem[Okamura and Yamada(2020)]{Okamura2020}
Kazuo Okamura and Seiji Yamada.
\newblock Adaptive trust calibration for human-{AI} collaboration.
\newblock \emph{PLOS ONE}, 15\penalty0 (2):\penalty0 e0229132, 2020.
\newblock \doi{10.1371/journal.pone.0229132}.

\bibitem[Zamfirescu-Pereira et~al.(2023)Zamfirescu-Pereira, Wong, Hartmann, et~al.]{ZamfirescuPereira2023}
J.~D. Zamfirescu-Pereira, Richmond~Y. Wong, Bjeorn Hartmann, et~al.
\newblock Why johnny can't prompt: How non-{AI} experts try (and fail) to design {LLM} prompts.
\newblock In \emph{Proceedings of the 2023 CHI Conference on Human Factors in Computing Systems}, CHI '23, pages 1--21, 2023.
\newblock \doi{10.1145/3544548.3581388}.

\bibitem[Lee and See(2004)]{Lee2004}
John~D. Lee and Katrina~A. See.
\newblock Trust in automation: Designing for appropriate reliance.
\newblock \emph{Human Factors: The Journal of Human Factors and Ergonomics Society}, 46\penalty0 (1):\penalty0 50--80, 2004.
\newblock \doi{10.1518/hfes.46.1.50_30392}.

\bibitem[Papenmeier et~al.(2019)Papenmeier, Englebienne, and Seifert]{Papenmeier2019}
Andrea Papenmeier, Gwenn Englebienne, and Christin Seifert.
\newblock How model accuracy and explanation fidelity influence user trust.
\newblock \emph{arXiv}, 2019.
\newblock \doi{10.48550/arXiv.1907.12652}.

\bibitem[Henrich(2020)]{Henrich2010}
Joseph Henrich.
\newblock \emph{The {WEIRDest} People in the World: How the West Became Psychologically Peculiar and Particularly Prosperous}.
\newblock Harvard University Press, 2020.

\bibitem[Seaborn et~al.(2023)Seaborn, Barbareschi, and Chandra]{Seaborn2023}
Katie Seaborn, Giulia Barbareschi, and Shruti Chandra.
\newblock Not only {WEIRD} but ``{U}ncanny''? {A} systematic review of diversity in human-robot interaction research.
\newblock \emph{International Journal of Social Robotics}, 15\penalty0 (11):\penalty0 1841--1870, 2023.
\newblock \doi{10.1007/s12369-023-01036-7}.

\bibitem[Tao et~al.(2024)Tao, Viberg, Baker, et~al.]{Tao2024}
Yan Tao, Olga Viberg, Ryan~S Baker, et~al.
\newblock Cultural bias and cultural alignment of large language models.
\newblock \emph{PNAS Nexus}, 3\penalty0 (9):\penalty0 346, 2024.
\newblock \doi{10.1093/pnasnexus/pgae346}.

\bibitem[Naous and Xu(2025)]{Naous2025}
Tarek Naous and Wei Xu.
\newblock On the origin of cultural biases in language models: From pre-training data to linguistic phenomena.
\newblock In \emph{Proceedings of the 2025 Conference of the Nations of the Americas Chapter of the Association for Computational Linguistics: Human Language Technologies (Volume 1: Long Papers)}, pages 6423--6443, 2025.
\newblock \doi{10.18653/v1/2025.naacl-long.326}.

\bibitem[Kharchenko et~al.(2024)Kharchenko, Roosta, Chadha, et~al.]{Kharchenko2024}
Julia Kharchenko, Tanya Roosta, Aman Chadha, et~al.
\newblock How well do {LLMs} represent values across cultures? empirical analysis of {LLM} responses based on hofstede cultural dimensions.
\newblock \emph{arXiv}, 2024.
\newblock \doi{10.48550/arXiv.2406.14805}.

\bibitem[Myung et~al.(2024)Myung, Lee, Zhou, et~al.]{BLEND2024}
Junho Myung, Nayeon Lee, Yi~Zhou, et~al.
\newblock {BLEnD}: Benchmark for everyday knowledge in diverse cultures.
\newblock In \emph{arXiv}, 2024.
\newblock \doi{10.48550/arXiv.2406.09948}.

\bibitem[Marchisio et~al.(2024)Marchisio, Ko, Berard, et~al.]{Marchisio2024}
Kelly Marchisio, Wei-Yin Ko, Alexandre Berard, et~al.
\newblock Understanding and mitigating language confusion in {LLMs}.
\newblock In \emph{Proceedings of the 2024 Conference on Empirical Methods in Natural Language Processing}, EMNLP 2024, pages 6653--6677, 2024.
\newblock \doi{10.18653/v1/2024.emnlp-main.380}.

\bibitem[Sambasivan et~al.(2021)Sambasivan, Arnesen, Hutchinson, et~al.]{Sambasivan2021}
Nithya Sambasivan, Erin Arnesen, Ben Hutchinson, et~al.
\newblock Re-imagining algorithmic fairness in india and beyond.
\newblock In \emph{Proceedings of the 2021 CHI Conference on Human Factors in Computing Systems}, CHI '21, 2021.
\newblock \doi{10.1145/3411764.3445518}.

\bibitem[Irani et~al.(2010)Irani, Vertesi, Dourish, et~al.]{Irani2010}
Lilly Irani, Janet Vertesi, Paul Dourish, et~al.
\newblock Postcolonial computing: A lens on design and development.
\newblock In \emph{Proceedings of the SIGCHI Conference on Human Factors in Computing Systems}, CHI '10, pages 1311--1320, 2010.
\newblock \doi{10.1145/1753326.1753522}.

\bibitem[Dourish and Mainwaring(2012)]{Dourish2012}
Paul Dourish and Scott~D. Mainwaring.
\newblock Ubicomp's colonial impulse.
\newblock In \emph{Proceedings of the 2012 ACM Conference on Ubiquitous Computing}, UbiComp '12, pages 133--142, 2012.
\newblock \doi{10.1145/2370216.2370238}.

\bibitem[Couldry and Mejias(2019)]{Couldry2019}
Nick Couldry and Ulises~A. Mejias.
\newblock \emph{The Costs of Connection: How Data Is Colonizing Human Life and Appropriating It for Capitalism}.
\newblock Stanford University Press, Stanford, CA, 2019.

\bibitem[Graham et~al.(2014)Graham, Hogan, Straumann, et~al.]{Graham2014}
Mark Graham, Bernie Hogan, Ralph~K. Straumann, et~al.
\newblock Uneven geographies of user-generated information: Patterns of increasing informational poverty.
\newblock \emph{Annals of the Association of American Geographers}, 104\penalty0 (4):\penalty0 746--764, 2014.
\newblock \doi{10.1080/00045608.2014.910087}.

\bibitem[Bignotti(2025)]{Bignotti2025}
francesca Bignotti.
\newblock Potential bias in ai: Cultural representation and the marginalization of african art.
\newblock In \emph{Proceedings of the AIUCD 2025 Conference}, AIUCD 2025, 2025.
\newblock URL \url{https://aiucd2025.dlls.univr.it/assets/pdf/papers/7.pdf}.

\bibitem[Murungu(2024)]{Murungu2024}
Ronnie Murungu.
\newblock Reimagining education in africa: The transformative potential of prompt engineering.
\newblock In \emph{OIDA International Journal of Sustainable Development}, 2024.
\newblock URL \url{https://oidaijsd.com/?page_id=3086}.

\bibitem[Asseri et~al.(2025)Asseri, Abdelaziz, and Al-Wabil]{Asseri2025}
Bushra Asseri, Estabrag Abdelaziz, and Areej Al-Wabil.
\newblock Prompt engineering techniques for mitigating cultural bias against arabs and muslims in large language models: A systematic review.
\newblock \emph{arXiv}, 2025.
\newblock \doi{10.48550/arXiv.2506.18199}.

\bibitem[Brown et~al.(2020)Brown, Mann, Ryder, et~al.]{Brown2020}
Tom Brown, Benjamin Mann, Nick Ryder, et~al.
\newblock Language models are few-shot learners.
\newblock In \emph{Advances in Neural Information Processing Systems}, NeurIPS 2020, pages 1877--1901, 2020.
\newblock URL \url{https://proceedings.neurips.cc/paper_files/paper/2020/file/1457c0d6bfcb4967418bfb8ac142f64a-Paper.pdf}.

\bibitem[Alhanai et~al.(2025)Alhanai, Kasumovic, Ghassemi, et~al.]{Alhanai2025}
Tuka Alhanai, Adam Kasumovic, Mohammad Ghassemi, et~al.
\newblock Bridging the gap: Enhancing {LLM} performance for low-resource african languages with new benchmarks, fine-tuning, and cultural adjustments.
\newblock In \emph{Proceedings of the 39th AAAI Conference on Artificial Intelligence}, AAAI 2025, pages 1--9, 2025.
\newblock \doi{10.48550/arXiv.2412.12417}.

\bibitem[Nakatumba-Nabende et~al.(2024)Nakatumba-Nabende, Babirye, Nabende, et~al.]{NakatumbaNabende2024}
Joyce Nakatumba-Nabende, Claire Babirye, Peter Nabende, et~al.
\newblock Building text and speech benchmark datasets and models for low-resourced {East African} languages: Experiences and lessons.
\newblock \emph{Applied AI Letters}, 5\penalty0 (2):\penalty0 e92, 2024.
\newblock \doi{10.1002/ail2.92}.

\bibitem[Graham et~al.(2011)Graham, Hale, and Stephens]{Graham2011}
Mark Graham, Scott~A. Hale, and Monica Stephens.
\newblock \emph{Geographies of the World's Knowledge}.
\newblock Convoco! Edition, London, 2011.

\bibitem[Pava et~al.(2025)Pava, Meinhardt, Uz~Zaman, et~al.]{Pava2025}
Juan Pava, Caroline Meinhardt, Haifa~Badi Uz~Zaman, et~al.
\newblock Mind the {L}anguage {G}ap: {M}apping the {C}hallenges of {LLM} {D}evelopment in {L}ow-{R}esource {L}anguage {C}ontexts.
\newblock Technical report, Stanford University, Institute for Human-Centered AI (HAI), April 2025.
\newblock URL \url{https://hai.stanford.edu/policy/mind-the-language-gap-mapping-the-challenges-of-llm-development-in-low-resource-language-contexts}.

\bibitem[Manvi et~al.(2024)Manvi, Khanna, Burke, et~al.]{Manvi2024}
Rohin Manvi, Samar Khanna, Marshall Burke, et~al.
\newblock Large language models are geographically biased.
\newblock \emph{Proceedings of the International Conference on Machine Learning (ICML)}, 2024.
\newblock \doi{10.5555/3692070.3693479}.

\bibitem[Adelani et~al.(2021)Adelani, Abbott, Neubig, et~al.]{Adelani2021}
David~Ifeoluwa Adelani, Jade Abbott, Graham Neubig, et~al.
\newblock Masakhaner: Named entity recognition for {A}frican languages.
\newblock \emph{Transactions of the Association for Computational Linguistics}, 9:\penalty0 1116--1131, 2021.
\newblock \doi{10.1162/tacl_a_00416}.

\bibitem[Talat et~al.(2022)Talat, Névéol, Biderman, et~al.]{Talat2022}
Zeerak Talat, Aurélie Névéol, Stella Biderman, et~al.
\newblock You reap what you sow: On the challenges of bias evaluation under multilingual settings.
\newblock In \emph{Proceedings of BigScience Episode \#5 -- Workshop on Challenges \& Perspectives in Creating Large Language Models}, pages 26--41, 2022.
\newblock \doi{10.18653/v1/2022.bigscience-1.3}.

\bibitem[Lee and Rich(2021)]{Lee2021}
Min~Kyung Lee and Katherine Rich.
\newblock Who is included in human perceptions of {AI}?: Trust and perceived fairness around healthcare {AI} and cultural mistrust.
\newblock In \emph{Proceedings of the 2021 CHI Conference on Human Factors in Computing Systems}, CHI '21, pages 1--14, May 2021.
\newblock \doi{10.1145/3411764.3445570}.

\bibitem[Afroogh et~al.(2024)Afroogh, Akbari, Malone, et~al.]{Afroogh2024}
Saleh Afroogh, Ali Akbari, Evan Malone, et~al.
\newblock Trust in {AI}: Progress, challenges, and future directions.
\newblock \emph{Humanities and Social Sciences Communications}, 11\penalty0 (1):\penalty0 1--30, 2024.
\newblock \doi{10.1057/s41599-024-03801-4}.

\bibitem[Dur\'{a}n and Pozzi(2025)]{Durán2025}
Juan~Manuel Dur\'{a}n and Giorgia Pozzi.
\newblock Trust and trustworthiness in {AI}.
\newblock \emph{Philosophy \& Technology}, 38\penalty0 (1):\penalty0 1--31, 2025.
\newblock \doi{10.1007/s13347-025-00843-2}.

\bibitem[Mehrotra et~al.(2024)Mehrotra, Degachi, Vereschak, et~al.]{Mehrotra2024}
Siddharth Mehrotra, Chadha Degachi, Oleksandra Vereschak, et~al.
\newblock A systematic review on fostering appropriate trust in human--{AI} interaction: Trends, opportunities and challenges.
\newblock \emph{ACM Journal on Responsible Computing}, 1\penalty0 (4):\penalty0 1--45, 2024.
\newblock \doi{10.1145/3696449}.

\bibitem[Jones et~al.(2023)Jones, Thornton, and Wyatt]{Jones2023}
Caroline Jones, James Thornton, and Jeremy~C. Wyatt.
\newblock Artificial intelligence and clinical decision support: clinicians’ perspectives on trust, trustworthiness, and liability.
\newblock \emph{Medical Law Review}, 31\penalty0 (4):\penalty0 501--520, 2023.
\newblock \doi{10.1093/medlaw/fwad013}.

\bibitem[Turner et~al.(2024)Turner, Kaushik, Huang, et~al.]{Turner2024}
Amy Turner, Meena Kaushik, Mu-Ti Huang, et~al.
\newblock Calibrating trust in {AI}-assisted decision making.
\newblock Technical report, UC Berkeley School of Information, 2024.

\bibitem[Coelho et~al.(2025)Coelho, Mirza, Cui, et~al.]{Coelho2025}
Bruno Coelho, Shujaat Mirza, Yuyuan Cui, et~al.
\newblock Understanding inequality of {LLM} fact-checking over geographic regions with agent and retrieval models.
\newblock \emph{arXiv}, 2025.
\newblock \doi{10.48550/arXiv.2503.22877}.

\bibitem[Azaroual(2024)]{Azaroual2024}
Fahd Azaroual.
\newblock Artificial intelligence in africa: Challenges and opportunities.
\newblock Technical Report PB-23/24, Policy Center for the New South, May 2024.

\bibitem[Gillespie et~al.(2023)Gillespie, Lockey, Curtis, et~al.]{Gillespie2023}
Nicole Gillespie, Steven Lockey, Caitlin Curtis, et~al.
\newblock Trust in artificial intelligence: A global study.
\newblock Technical report, The University of Queensland and KPMG Australia, 2023.

\bibitem[{Republic of Kenya, Ministry of Information, Communications, and Digital Economy}(2025)]{KenyaAI2025}
{Republic of Kenya, Ministry of Information, Communications, and Digital Economy}.
\newblock Kenya's artificial intelligence strategy 2025-2030.
\newblock Government Publication, 2025.
\newblock URL \url{https://ict.go.ke/sites/default/files/2025-03/Kenya%20AI%20Strategy%202025%20-%202030.pdf}.

\bibitem[Federal Republic~of Nigeria(2025)]{NigeriaAI2025}
Office of the National Security~Adviser Federal Republic~of Nigeria.
\newblock National artificial intelligence strategy (nais) 2025.
\newblock Government Publication, 2025.
\newblock URL \url{https://ncair.nitda.gov.ng/wp-content/uploads/2025/09/National-Artificial-Intelligence-Strategy-19092025.pdf}.

\bibitem[Van~de Mortel(2008)]{VanDeMortel2008}
Thea~F. Van~de Mortel.
\newblock Faking it: Social desirability response bias in self-report research.
\newblock \emph{Australian Journal of Advanced Nursing}, 25\penalty0 (4):\penalty0 40--48, 2008.
\newblock URL \url{https://www.ajan.com.au/archive/Vol25/Vol_25-4_vandeMortel.pdf}.
\newblock ISSN 0813-0531.

\end{thebibliography}

\end{document}